\font\twelvebf=cmbx12
\font\ninerm=cmr9
\nopagenumbers
\magnification =\magstep 1
\overfullrule=0pt
\baselineskip=18pt
\line{\hfil }
\line{\hfil December 1997}
\vskip .8in
\centerline{\twelvebf  Infinite Statistics and Black Holes in Matrix Theory}
\vskip .5in
\centerline{\ninerm  D.MINIC}
\vskip .1in
\centerline{Physics Department}
\centerline{Pennsylvania State University}
\centerline{University Park, PA 16802}
\vskip .1in
\centerline {
minic@phys.psu.edu}

\vskip 1in
\baselineskip=16pt
\centerline{\bf Abstract}
\vskip .1in

The concept of infinite statistics is applied to the analysis of
black hole thermodynamics in Matrix Theory.
It is argued that Matrix Theory partons, D0-branes, satisfy  
quantum infinite statistics, and that this observation  
justifies the recently proposed Boltzmann gas model of
Schwarzschild black holes in Matrix Theory.

\vfill\eject

\footline={\hss\tenrm\folio\hss}

\magnification =\magstep 1
\overfullrule=0pt
\baselineskip=22pt
\pageno=1

{\bf 1.} Various authors have recently shown that Matrix theory [1] can be
used as a starting point for the analysis of basic qualitative
features of Schwarzschild black holes [2], [3], [4], [5], [6]. (For
other approaches  to the same subject consult [7].) 
According to [2], [3], [4], [5], [6], neutral black holes can be understood
as bound states of the zero modes of the induced super Yang-Mills
theory which appears in the framework of
a toroidally compactified Matrix Theory.
For example, the Bekenstein-Hawking entropy formula
 can be deduced by analyzing 
the mean field theory  
 of the induced super Yang-Mills 
zero modes. 
 The fundamental relation between the black hole entropy and its  
mass then follows from the defining dispersion relation of the infinite
momentum frame kinematics (one of the salient properties of Matrix theory).
The original analysis of [2], [3] and [4] was done in 
a particular limit defined by the condition that
 the entropy $S$ of a neutral 
black hole configuration is 
proportional to the number of Matrix theory partons $N$.
(The other important ingredient was the use of particular
equations of state for $d +1$ super Yang-Mills theories with
16 supercharges, derived in the study of the thermodynamics of
near-extremal D-branes [15], [16].) 
It was further 
elucidated in [5] that the 
$N \sim S$  limit arises at a black hole/black
string transition point,  
at which longitudinally stretched black strings become unstable to
the formation of black holes. 
In this case the thermodynamics of neutral black holes, it was
 argued in [5], is completely determined by the 
mean field dynamics of the 
induced  
super Yang-Mills zero modes. Actually, the limit $N \gg S$ 
is in some sense 
much more appropriate for the infinite momentum frame black hole 
physics, for in
this situation, as pointed out in [6], the black hole entropy is 
independent from the specifics of the boosting procedure. As it turns out,
in both cases 
($N\sim S$ and $N \gg S$) 
the fundamental properties of infinitely boosted neutral black holes can be
understood in the framework of an interacting gas of Matrix theory partons,
the nature of interactions being dictated by the fundamental Matrix
theory Hamiltonian [3], [5], [6].

It was emphasized in [3] that 
 Matrix theory partons, i.e. D0-branes [23], [24], [25], [26], 
should be treated as distinguishable particles;  
otherwise the basic relations of black hole thermodynamics
would not follow. According to [3], D0-branes are essentially 
 distinguishable (Boltzmannian) because they can
be understood as fluctuations around particular classical 
background configurations of Matrix theory (in the language of
[3] D0-branes  
 are tethered to  particular classical background configurations).
Different classical backgrounds which are 
responsible for the Boltzmannian character of 
D0-branes have to be 
 identified in different dimensions [3]. 

More recently, the basic properties of a classical Boltzmann gas
of D0-branes were analyzed in [14]. The  
classical partition function was computed for an interacting Boltzmann
gas of D0-branes
whose dynamics was modeled by a Born-Infeld type of lagrangian
 [17], [19], [23], [26].
It was pointed out in [14] that the absence of the Gibbs $1/N!$ factor
($N$ being the number of distinguishable particles) in the 
classical Boltzmann partition function is essential for computation
of black hole thermodynamic functions, thus confirming the
intuition of [3]. 

It is our aim in this article to further elaborate on
this issue. We want to point out that the absence of the Gibbs factor and
the Boltzmann nature of D0-branes follow from the fact
that D0-branes obey quantum
infinite statistics [9], [10]. Unlike the familiar Bose-Einstein
or Fermi-Dirac statistics, somewhat exotic infinite statistics allow
 all representations of the
symmetric group  to be realized.  
Because of this fact, various quantum statistical properties of
many-body systems described by infinite statistics are very similar
to the corresponding properties of classical Boltzmann systems.
Hence the intuition of [3] and the results of [14] are
justified from a quantum statistical point of view.
The resulting picture also turns out to be in 
full agreement with the ideas of Strominger and Volovich [8] concerning
black hole statistics. 

The note is organized as follows: First we review the
basic set-up of the recently proposed 
Matrix theory description of Schwarzschild 
black holes as presented in [2], [3], [4], [5], [6].
 Then we review the concept of infinite statistics
following the original work of [9] and apply it initially to a gas of
free particles. Next we consider a
general non-ideal gas of particles described by infinite statistics.
We establish constraints that have to be satisfied by the 
effective Hamiltonian which describes the dynamics of
 such a gas in order for it to 
qualify as a model 
 of black hole thermodynamics in $D$ space-time dimensions ($D > 4$). 
The resulting
effective Hamiltonian is found to agree with Matrix theory predictions. 
Our basic point is
that quantum infinite statistics of D0-branes justifies the
intuitive picture of the Boltzmann gas model of Schwarzschild
black holes in Matrix theory [3] and explains the success of
 calculations performed in [14].

{\bf 2.} Let us review the basic physics of the Matrix theory approach to
Schwarzschild black holes [2], [3], [5], [6].
We consider Matrix theory compactified on a $d$-dimensional
torus $T^{d}$, the 
dimension of the uncompactified space time being $D=11-d$ .
Let the characteristic size of the torus $T^{d}$ be $L$, and
the extent of the compact M-theory dimension be $R$ (where $R$,
is eventually taken to infinity).
  
The entropy of a neutral black hole in $D$ space-time dimensions
is proportional to the transverse spatial volume and
 is given by the Bekenstein-
Hawking formula
$$
S \sim R_{s}^{D-2}/G_{D}, \eqno(1)
$$
where $G_{D} = l_{p}^{9}/L^{d}$ is the $D$-dimensional Newton constant
and the Schwarzschild radius $R_{s}$ has the following dependence on the
black hole mass $M$ (as determined by the $r^{3-D}$ form of the
gravitational Newtonian potential in $D$ space-time dimensions) 
$$
R_{s} \sim (G_{D} M)^{1 \over {D-3}}.  \eqno(2)   
$$
The dynamics of black holes is studied in the infinite momentum
frame (IMF) (one of the defining features of
Matrix theory) which is characterized by the following fundamental 
kinematic dispersion relation between the light cone energy $E$ and
the longitudinal momentum $P \sim N/R$  
$$
E \sim M^{2}R/N.  \eqno(3)
$$
Equation (3) together with the first law of thermodynamics $dE =T dS$
 implies 
the following expression   
$$
S \sim (NT/R)^{{D-2} \over {D-4}} G_{D}^{2 \over {D-4}}, \eqno(4)
$$
or equivalently, 
$$
R_{s} \sim (NTG_{D}/R)^{1 \over {D-4}}.  \eqno(5)
$$

The above basic relations can be deduced from a  mean field analysis
of the dynamics of the zero modes of the induced super Yang-Mills theory
[3],[5],[6].
The argument goes as follows: The energy and longitudinal
momentum of a boosted neutral black hole, with the boosting
parameter $e^{\alpha}$, 
are 
$$
E \sim M \cosh{\alpha} ,  P \sim M \sinh{\alpha}.   \eqno(6)
$$
These kinematic relations determine
 (at the black hole/black string transition point $N \sim S$,
that is $P \sim S/R$)
by how much the box of size $R$ has to expand in order to accomodate
the black hole of size $R_{s}$, in the case of very large boosts, namely 
$$
\exp{\alpha} \sim R_{s}/R. \eqno(7)
$$

Assume that neutral black holes can be viewed  
as bound states of Matrix theory partons. It is well known that the  
mean field dynamics of $N$ zero modes
of the induced super Yang-Mills theory
 is governed by the following effective Matrix theory Lagrangian [1],
[24], [25] 
(derived in the limit of low parton velocities $v$ and large interparton 
separations
$r$)  
$$
L_{eff} = {{Nv^{2}} \over R} + G_{D} {{N^{2} v^{4}} \over {R^{3} r^{D-4}}}
. \eqno(8)  
$$ 
Suppose one applies the virial theorem 
(i.e. equate the average kinetic and potential energies)
to such a black hole-like  
bound state, by setting the characteristic size of the bound state
to be essentially the Schwarschild radius 
$R_{s}$ and the characteristic parton velocity
to be proportional to the inverse of the boosting parameter 
$v \sim e^{-\alpha} \ll 1 $ (or $vR_{s}/R \sim 1$, which is just the
Heisenberg uncertainty bound). Then one can easily
deduce the Bekenstein-Hawking
relation (1) at the special point $N \sim S$, in other words, 
find that $N \sim R_{s}^{D-2}/G_{D}$! The use of the IMF on-shell relation (3)
then implies $M \sim R_{s}^{D-3}$, which is identical to equation (2)!  
(Note that the first law of thermodynamics is not used in this
part of the argument.) 

On the other hand, if one examines the limit of $R \gg R_{s}$, one
is lead to consider the case $N \gg S$, which in turn
corresponds to the limit of low temperatures [6]. In this
particular situation one may again apply the virial theorem
in the following manner    
$$
N m <v^{2}> \sim E \sim TS.   \eqno(9)
$$
Here the characteristic mean velocity is determined as 
$v \sim R_{s} T$ ($R_s$ being the typical distance and
$T$- typical frequency of the system) 
and the parton mass $m \sim 1/R$. (Note that in this limit the first law of
thermodynamics is used.)  Then equation (9) implies 
$R_{S} \sim (NTG_{D}/R)^{1 \over {D-4}}$, which is identical
to equation (5)! Moreover, as shown in
[6], one can deduce the form of interactions needed
to derive the equation of state (4)
 and find that, indeed, the 
required interactions (which as it turns out
involve spin [27], [28], [29]) can be deduced from the fundamental
Matrix theory Hamiltonian. 

{\bf 3.} In this section we want to show that all fundamental relations
of section 2. follow from the quantum statistical mechanics of 
particles obeying
infinite statistics (D0-branes being such particles).

We start with a review of infinite statistics following the
original reference [9]. The construction
 of [9] represents a concrete realization
of the results of [10] in which a particular example of 
statistics allowing for 
all possible representations of the symmetric group, 
 was found (along with more familiar Bose-Einstein, Fermi-Dirac and 
parastatistics, the last being essentially equivalent  
to internal "color" symmetry ). The explicit operator realization of infinite
statistics in [9] 
is given in terms of the  
Cuntz algebra [11] (used, for example,
 in non-commutative probability theory [12] and
its applications to the study of the large $N$ ('t Hooft) limit of
matrix models [13]) 
$$
a_i a^{\dagger}_j = \delta_{ij}.  \eqno(10)  
$$
As usual, a unique vacuum state is assumed. The vacuum state $|0\rangle$
is  defined via
$$
a_i |0\rangle =0.              \eqno(11)  
$$
Notice that the infinite statistics commutation relations 
(10) can be thought of, rather formally, as a $q=0$ limit of the
$q$-deformed quantum
 commutation relations $a_i a^{\dagger}_j - q a^{\dagger}_j  
a_i = \delta_{ij}$ which describe "q-ons" [9]. 

Many unusual properties are implied
by (10). For example, the number
operator $N_i$ ( $[N_i, a_j] \equiv - \delta_{ij} a_j$) is given by
the following expression 
$$
N_{i} = a_i^{\dagger} a_i + \sum_k a_k^{\dagger} a_i^{\dagger}
a_i a_k + \sum_{k_1, k_2} a_{k_1}^{\dagger} a_{k_2}^{\dagger}
a_i^{\dagger} a_i a_{k_2} a_{k_1} + ... \eqno(12)  
$$
Also, the inner product of two $n$-particle states is determined by
$$
<0|a_{i_n}...a_{i_1}a^{\dagger}_{j_1}...a^{\dagger}_{j_n}|0> =
\delta_{i_1 j_1}...\delta_{i_n j_n} .   \eqno(12a)  
$$
Similarly, the 
operators which create (annihilate) a particle in a given quantum state
$k$ read as follows 
$$
A^{\dagger}_k (A_k) = \sum_{i_1,...,i_l} a^{\dagger}_{i_1}...a^{\dagger}_{i_l}
a^{\dagger}_{k} (a_k) a_{i_l}...a_{i_1} .   \eqno(12b)
$$
Because infinite statistics particles belong to many-dimensional
representations of the symmetric group, the place in which these
operators act is important.

It was argued in [9] that there exists no 
second quantized formulation of a local field
theory in terms of such "free" operators (i.e. operators obeying (10)),  
  yet there exist consistent non-relativistic theories.  
Moreover, the spin-statistics theorem implies no restriction on spin for
particles satisfying infinite statistics. Furthermore, the principle
of cluster decomposition and CPT theorem are found to hold [9]. 

We are interested in the quantum statistical properties of 
many-body systems
described by infinite statistics. More specifically,
 we want to argue that a gas of
$N$ ($N\rightarrow \infty$) D0-branes 
behaves effectively as such a 
 many-body system.  
 The basic object of our study is the
quantum partition function which is
defined as usual  
$$
Z \equiv \sum e^{- \beta H},     \eqno(13) 
$$
where the sum goes over all possible quantum states. 

Let us first consider a free gas of particles
satisfying infinite statistics.
For such a free gas one 
can easily see that the quantum partition function (13)
has the form of the 
  classical Boltzmann partition function without the 
Gibbs $1/N!$ factor [9]. Put differently, the quantum partition function of
a free gas of infinite statistics particles effectively describes
the statistical properties of a system of
 identical particles with a very large (infinite) 
number of internal degrees of freedom, so that particles can be
distinguished by their internal quantum numbers. In the case of
interest, that is, a free gas of $N$ ($N \rightarrow \infty$)
 D0-branes, the particular internal symmetry is $U(\infty)$ [1].
Our basic argument is simple:
free D0-branes can be distinguished by their internal states and are
thus effectively described by infinite statistics.
Notice that this fact is the 
underlying reason why there exists no Gibbs paradox 
in this situation (which, we recall, 
is the generic feature of classical Boltzmann gases). 

The explicit form of the
quantum 
partition function for a free gas of particles obeying infinite
statistics (in $D-2$ transverse dimensions) is   
$$
Z \sim V^N {(T/R)}^{N {{D-2} \over 2}}.      \eqno(14)  
$$
Here $V$ denotes the volume of the
transverse space and the parton mass $m \sim 1/R$. 
 Notice how this differs from the usual expression for the
classical partition function (which includes the
Gibbs factor) by a 
factor $(1/N)^N$.

The energy of an ideal infinite statistics gas is therefore given by
$$
E = - {{\partial \ln{Z}} \over {\partial \beta}} \sim {N \over \beta} 
, \eqno(15)
$$
and its entropy
$$
S = \ln{Z} + \beta E \sim \ln{Z} + N .  \eqno(16) 
$$
Now, if we assume that $\ln{Z} \ll N$ i.e. $Z \sim 1$,
we get that the entropy is proportional to the
number of infinite statistics particles $S \sim N$! The requirement
$Z \sim 1$ (which is "optimal" from the renormalization group 
point of view, like the condition $N \sim S$ [2], [5]) amounts to
the condition
$$
V (T/R)^{{D-2} \over 2} \sim 1.  \eqno(16a)  
$$
What is the meaning of this relation?
It is easy to see that,
for $N \sim S$, equations (1), (2) and (3) (without using
the first law of thermodynamics) lead to 
$$
S \sim (T/R)^{-{{{D-2} \over 2}}} ,  \eqno(16b)
$$
which is compatible with (16a) once the Bekenstein-Hawking
formula (1) is recalled!   
In that sense the requirement $Z \sim 1$ is just another
way of saying that the holographic principle [1], [21], [22] is
at work here.

Let us now study a non-ideal, i.e. interacting gas of particles
obeying infinite statistics.  
We are interested in the post-Newtonian limit (small velocities $v$,
large separations $r$ between particles) which is quite natural from the
point of view of the dynamics of D0-branes in Matrix
theory [1], [23], [24], [25].
The most general form of such a mean field post-Newtonian lagrangian (if for
the moment we imagine possible dependence on spin, so that
time-reversal invariance is not violated) is
$$
L \sim \sum_{l=0, k=0} a_{lk} v^{l+2} r^{-k},    \eqno(17)  
$$
where we have expanded around the quadratic kinetic term  which
describes the ideal gas, so that $a_{00} \sim N$.
In general the partition function (13) cannot be evaluated in
a closed form. But if the interaction potential is of the order of the
kinetic term, the quantum partition function can be readily estimated.
In fact, the quantum partition function reduces to equation (14)
in this case. 
(Notice that our notation is quite schematic; the explicit sums over
particles as well as indices determining particle positions have
been suppressed for simplicity).

We would ultimately like to describe a Schwarzschild black hole
as a bound state of interacting infinite statistics particles.
Hence we apply the virial theorem to the lagrangian (17) 
and demand that the $i$-body interaction is of the
same order as the $i+j$-body interaction and that
 the Bekenstein-Hawking relation (1) remains valid
for $N \sim S$. The end product
is the
following mean field lagrangian 
$$
L \sim \sum_{i} b_{i} v^2 (N v^{m} r^{-D+m+2})^{i}.   \eqno(17a)
$$
Note that here the general
$i$-body interaction goes as $N^i$, the factor which replaces
explicit sums over interacting particles (also, $b_0 \sim N$).
We naturally assume, as in section 2.,
 that the characteristic size of the bound state
is of the order of the Schwarzschild radius $R_s$ and
that the characteristic velocities of particles saturate the Heisenberg
uncertainty bound. The virial theorem demands then that
$N v^{m} \sim r^{D-m-2} $, which exactly corresponds to the
Bekenstein-Hawking entropy formula (1), for $N \sim S$.    
The quantum partition function is determined by (14),
which is also consistent with the fact that $N \sim S$. 

Note that in case we wish to describe a gravity-like interaction in
$D$ space-time dimensions (and in the infinite momentum
limit), the static potential is uniquely determined 
to be the static transverse-space Newtonian potential 
i.e. $r^{-D+4}$ which immediately implies $m=2$ in (17a).

The question arises: How does Matrix theory predictions fit into
this particular result?

To start with, the holographic nature of the 
fundamental Hamiltonian of Matrix theory [1], [21], [22]   
determines 
the most general form of the effective post-Newtonian 
lagrangian (in case we neglect spin dependence) 
[17], [18], [19],
[20]  
$$
L_n \sim \sum_{m=0}^{\infty} c_{nm} {({N v^2 \over R})}
{({N v^2 \over r^{D-4}})}^n {({v^2 \over r^4})}^{m-n} 
{({1 \over R^2})}^m ,   \eqno(18)   
$$
($n$ denotes the loop order). This lagrangian
agrees with the form of (17). In the special case when $n=m$
$$
L_{diag} = \sum_{n} c_{nn} {({N v^2 \over R})} 
{({ N v^2 \over {R^2 r^{D-4}}}
)}^n .  \eqno(18a)
$$
For example, the effective lagrangian
 (8) represents the first two terms from (18a).
It is again easily seen that (18a) agrees with the form of (17a).

Note that
equation (18a) is also in full agreement with the post-Newtonian expansion 
 of
a $D$-dimensional 
 Einstein-Hilbert action [17], [18], [19]. 
 As a matter of fact (18a) represents
the expansion of a Born-Infeld type of lagrangian considered in [14],
 [17]
(with fixed coefficients $c_{nn}$ [18])
$$
L_{B-I} \sim h^{-1} (\sqrt{1 - h v^{2}} -1),  \eqno(18b)
$$
and with $h \sim r^{-(D-4)}$. 
In particular, the classical statistical mechanics of such an effective
lagrangian without the Gibbs $1/N!$ term is studied in [14]. The classical
partition function turns out to be, 
perhaps not surprisingly, equivalent to (14). 

To recapitulate: The virial theorem (when applied to (18a) or
equivalently (18b))
implies that $N{v^2 \over r^{D-4}} \sim 1$, so that the
quantum partition function in this approximation becomes
(14), as it should. In addition, the
requirement that $r^{D-4} \sim N v^2$ once again corresponds to
the Bekenstein-Hawking formula (1), the moment we apply the
same estimates for the velocity and distance as in section
2.

 Note that all this indicates that
 a neutral black hole-like bound state of infinite
statistics particles cannot be   
described by  the full effective lagrangian (18), even
though (18) is derived from a holographic theory.
This particular fact 
is in accordance with the physical picture developed
in [6] which states that a neutral black hole
bound state is determined by a highly coherent set of interactions
which are all of the same strength.
The form of the selected interactions follows from the effective
lagrangian (18a,b) or, in other words, from a post-Newtonian expansion
of the leading term in the low energy action of Matrix theory, namely, 
the $D$ dimensional Einstein-Hilbert term .  
 
The above analysis applies to the situation when $Z \sim 1$ or, as
 described by 
 eq. (16) when $N \sim S$. What happens in the case when
$N \gg S$? The result of [6] should apply then. From this
reference we read off the form of
the necessary $l$-body effective interactions   
$$
U_{eff} \sim ({N/R})^l {v^{l+1} \over r^{(l-1)(D-3)}}, \eqno(19)
$$
where we have dropped the explicit spin dependence which is
needed for time-reversal invariance.

Let us apply the virial theorem to this physical situation. The
interaction terms are of the order of the kinetic term, and therefore, 
$N {v \over r^{D-3}} \sim R$ .
 This requirement  leads (in the
limit $N \gg S$ ) 
to the correct dependence of the Schwarzschild radius $R_s$
on temperature (when, as reviewed in section 2. $v \sim R_s T$), i.e.  eq. (5). Note that if
we consider the $N \sim S$ limit (when $v \sim 1/R_{s}$), 
the same interactions (19)
would turn out to be compatible with the 
general expression (17a). The requirement
$N {v \over r^{D-3}} \sim R$  is then equivalent to the
Bekenstein-Hawking relation (1), as expected.

To summarize: Schwarzschild black holes can be viewed as bound
states of particles obeying quantum infinite statistics (such as
D0-branes)
if the effective
interaction between these particles is given by the "diagonal"
(Einstein-Hilbert or Born-Infeld) 
part of the Matrix theory effective lagrangian (18a). 
This interacting D0-brane gas is of Boltzmann nature [3], [14],
because of the effective quantum infinite statistics of its constituents.
The entropy of such a gas is proportional to the number of
infinite statistics particles, when the value of the quantum
partition function is of the order of unity, in accordance with
the holographic principle of infinite momentum frame black hole
dynamics [21], [22].  

The resulting
 picture nicely meshes with the original ideas of Strominger and
Volovich [8] concerning the issue of black hole statistics. 
Strominger suggested that the quantum statistics of
extremaly charged four-dimensional black holes is neither Bose-Einstein
nor Fermi-Dirac, but rather infinite statistics. Strominger assumed
that the black hole quantum state is a functional on the space of closed 
three-geometries, with each black hole connected to an oppositely charged
black hole via a spatial wormhole. By examining the process of black
hole exchange, which is the exchange of two wormhole ends, and which
generically results in a new three-geometry, Strominger concluded that
the wave function does not have any specific symmetry properties under
this operation; the wave function for many extremal black holes is
a general function of black hole positions. Hence, 
black holes resemble distinguishable particles. (Volovich suggested an
extension of this
idea to D-branes, given the similarity of black holes and D-branes.)

{\bf 4.} In conclusion, in this note we have argued that quantum
infinite statistics of D0-branes is what underlies   
 the recently proposed Boltzmann gas model of Schwarzschild
black holes in Matrix theory [3], [14]. It would be very
interesting to apply the outlined formalism of 
infinite statistics [9] to the problem of Hawking radiation.
For example, the form of creation/annihilation operators as given
by (12b), is very suggestive when thinking about the analogy  
between D0-branes and Hawking particles. The obvious subtle point
is that Hawking radiation can be thought of as a $1/N$ effect,
whereas the quantum infinite statistics commutation relation (10) nicely
fits the usual large $N$ framework [12], [13].
We can only speculate about the possibility
 that the general $q$-on algebra
$a_i a^{\dagger}_j - q a^{\dagger}_j a_i = \delta_{ij}$ with
$q \sim 1/N$ might turn out to be useful for the Hawking radiation problem.

\vskip .1in
{\bf Acknowledgements}

I would like to thank Shyamoli Chaudhuri and Miao Li for many 
discussions on the topic of black holes in Matrix Theory.


\vskip.1in
{\bf References}
\vskip .1in
\item{1.} T. Banks, W. Fischler, S. H. Shenker and L. Susskind, Phys. 
Rev. D55
(1997) 5112. For a review and further references, see T.Banks, 
hep-th/9706168; D. Brigatti and L. Susskind, hep-th/9712092.
\item{2.} T. Banks, W. Fischler, I. Klebanov and L. Susskind,
hep-th/9709091 
\item{3.} T. Banks, W. Fischler, I. Klebanov and L. Susskind,
 hep-th/9711005.
\item{4.} I. Klebanov and L. Susskind, hep-th/9709108.
\item{5.} G. Horowitz and E. Martinec, hep-th/9710217.
\item{6.} M. Li, hep-th/9710226.
\item{7.} 
 M. Li and E. Martinec, hep-th/9703211, hep-the/9704134; R. Dijkgraaf,
E. Verlinde and H. Verlinde, hep-th/9704018; E. Halyo, hep-th/9705107.
\item{8.} A. Strominger, Phys. Rev. Lett. 71 (1993) 3397  
; I. V. Volovich, hep-th/9608137.
\item{9.} O.W. Greenberg, Phys. Rev. Lett. 64 (1990) 705; Phys. Rev.
D43 (1991) 4111; hep-ph/9306225.
\item{10} S. Doplicher, R. Haag, J. Roberts, Comm. Math. Phys. 23 (1971)
199; Comm. Math. Phys. 35 (1974) 49.
\item{11.} J. Cuntz, Comm. Math. Phys. 57 (1977) 173.
\item{12.} D. Voiculescu, K. Dykema and A. Nica
 {\it "Free Random Variables", American Mathematical
Society, 1992.}
\item{13.} M. R. Douglas, Phys. Lett. B344 (1995) 117; Nucl. Phys. Proc.
Suppl. 41 (1995) 66 ; M. R. Douglas and M. Li, Phys. Lett. B348 (1995)
360; R. Gopakumar and D. Gross, Nucl. Phys. B451 (1995) 379; A. Migdal, 
Nucl. Phys. Proc. Suppl. 41 (1995) 141;
 I. Aref'eva and I. V. Volovich, Nucl. Phys. B462 (1996) 600; D. Minic,
hep-th/9503203; hep-th/9502117; Some earlier references are:
O. Haan, Z. Phys. C6 (1980) 345; P. Cvetanovic, Phys. Lett. B99 (1981) 49;
A. Jevicki and H. Levine, Ann. Phys. 13 (1981) 61. M. B. Halpern and C.
Schwartz, Phys. Rev. D24 (1981) 2146.  
\item{14.} H. Liu, A. A. Tseytlin, hep-th/9712063.
\item{15.} S. S. Gubser, I. R. Klebanov and A. W. Peet, Phys. Rev. D54 (1996)
3915.
\item{16.} I. R. Klebanov and A. A. Tseytlin, Nucl. Phys. B475 (1996) 165.
\item{17.} K. Becker, M. Becker, J. Polchinski and A. A. Tseytlin,
Phys. Rev. D56 (1997) 3174.
\item{18.} K. Becker and M. Becker, hep-th/9705091.
\item{19.} E. Keski-Vakkuri and P. Kraus, hep-th/9709122; 
hep-th/9711013. 
\item{20.} P. Berglund and D. Minic, hep-th/9708063 .
\item{21.} G. 't Hooft, gr-qc/9310026.
\item{22.} L. Susskind, Jour. Math. Phys. 36 (1996) 6377  .
\item{23.} C. Bachas, Phys. Lett. B374 (1996) 37.
\item{24.} U. Danielsson, G. Ferretti and B. Sundborg, Int. J. Mod. Phys.
A11 (1996) 5463; D. Kabat and P. Pouliot, Phys. Rev. Lett. 77 (1996) 1004.
\item{25.} M. R. Douglas, D. Kabat, P. Pouliot and S. H. Shenker, Nucl. Phys.
B485 (1997) 85.
\item{26.} J. Polchinski, Phys. Rev. Lett. 75 (1995) 4724. J. Dai,
R. G. Leigh, J. Polchinski, Mod. Phys. Lett. A4 (1989) 2073;
R. G. Leigh, Mod. Phys. Lett. A4 (1989) 2767; J. Polchinski,
{\it TASI Lectures on D-branes}. 
\item{27.} H. Awata, S. Chaudhuri, M. Li and D. Minic, hep-th/9706   .
\item{28.} J. Harvey, hep-th/9706039.
\item{29.} P. Kraus, hep-th/9709199.

\end